\documentclass[12pt]{article}

\usepackage{amsmath,bm,graphicx,color,bm,color,soul,url,threeparttable,amsfonts,natbib}

\def\bSig\mathbf{\Sigma}

\newtheorem{lemma}{Lemma}
\newtheorem{theorem}{Theorem}
\newtheorem{remark}{Remark}

\def\T{{ \mathrm{\scriptscriptstyle T} }}

\def\pr{\mbox{pr}}
\def\bmTheta{\bm{\Theta}}

\title{\bf Efficient inference of parental origin effects using case-control mother-child genotype data}
\author{Yuang Tian$^1$, Hong Zhang$^{2}$, Alexandre Bureau$^3$, Hagit Hochner$^4$, Jinbo Chen$^{5}$}
\date{}

\begin{document}

\maketitle

\noindent $^1$Shanghai Center for Mathematical Sciences, Fudan University, Shanghai, China 

\noindent  $^2$Department of Statistics and Finance, School of Management, University of Science and Technology of China, Hefei, Anhui, China

\noindent  $^3$Department of Social and Preventive Medicine, Université Laval, Québec, Canada

\noindent  $^4$Braun School of Public Health, The Hebrew University of Jerusalem, Jerusalem, Israel 

\noindent  $^5$Department of Biostatistics and Epidemiology, Perelman School of Medicine, University of Pennsylvania, Philadelphia, PA, U.S.A. 

\newpage

\begin{abstract}
Parental origin effects play an important role in mammal development and disorder. Case-control mother-child pair genotype data can be used to detect parental origin effects and is often convenient to collect in practice. Most existing methods for assessing parental origin effects do not incorporate any covariates, which may be required to control for confounding factors. We propose to model the parental origin effects through a logistic regression model, with predictors including maternal and child genotypes, parental origins, and covariates. The parental origins may not be fully inferred from genotypes of a target genetic marker, so we propose to use genotypes of markers tightly linked to the target marker to increase inference efficiency. A computationally robust statistical inference procedure is developed based on a modified profile likelihood in a retrospective way.
A computationally feasible expectation-maximization algorithm is devised to estimate all unknown parameters involved in the modified profile likelihood. This algorithm differs from the conventional expectation-maximization algorithm in the sense that it is based on a modified instead of the original profile likelihood function. The convergence of the algorithm is established under some mild regularity conditions. This expectation-maximization algorithm also allows convenient handling of missing child genotypes. Large sample properties, including weak consistency, asymptotic normality, and asymptotic efficiency, are established for the proposed estimator under some mild regularity conditions. Finite sample properties are evaluated through extensive simulation studies and the application to a real dataset.

 
\end{abstract}

Running title: Inference of parental origin effects

\newpage

\section{Introduction}
\label{s:intro}

Parental origin effects exist when the genetic effects of alleles inherited from parents depend on their parental origin. Parental origin effects play an important role in certain neonatal diseases, such as Beckwith-Wiedemann, Prader-Willi, and Angelman syndrome \citep{falls1999genomic}. There is evidence that parental origin effects of fetal genes may also be involved in the development of maternal pregnancy-related diseases \citep{kanayama2002deficiency, petry2007does, saftlas2005immunogenetic, wangler2005factors}.

Most studies for testing parental origin effects used traditional family designs because detecting these effects requires familial information, including the transmission disequilibrium test \citep{weinberg1999methods,purcell2007plink,orton2011vitamin,wang2012ror2}. Due to high cost of recruiting father-mother-child trios \citep{yang2013robust}, it is an intermediate solution to use a mother-child pair design, which is widely used in the studies of neonatal diseases \citep{fu2013testing,mendoncca2019epigenetic,van2020maternal}. There have been several studies that used mother-child data to assess parental origin effects \citep{ainsworth2011investigation,weinberg2009genetics,howey2012premim} by applying a log-linear model to single nucleotide polymorphism (SNP) data. The major difficulty in studying parental origin effects is that the parental origin of the children alleles cannot be unambiguously determined when the genotypes of both mother and child are heterozygous. The power for testing parental origin effects could be substantially improved by exploiting linkage disequilibrium information between the target SNP and its adjacent SNPs since such information can assist to determine parental origin of child alleles \citep{browning2009unified,kong2009parental,howey2015increased,zhang2020incorporating}. Using mother-child paired genotype data, \cite{lin2013multi} developed a likelihood based efficient statistical inference method by fully utilizing the information from adjacent SNPs. Furthermore, some studies considered using data from case-parent trios \citep{weinberg1998log,cordell2004case} or other types of family structures \citep{han2013joint,zhang2016optimum,zhang2019imprinting}.

Some studies use discrete covariates for stratification \citep{gjessing2006case,gjerdevik2018parent}, but none of the above methods can adjust for continuous covariates in a general way. In this paper, we propose to model genetic effects through haplotypes related to SNPs tightly linked to the target SNP, adjusting for environmental factors. Considerable efficiency gain can be achieved since the information from adjacent SNPs is utilized. We derive a modified profile likelihood function for the observed data by treating the joint distribution of maternal genotype and covariates to be fully nonparametric, so that our method is robust to some extent. By treating the haplotypes associated with multiple SNPs as missing data, we extend the conventional expectation-maximization algorithm to maximize the modified profile likelihood function, and show that the algorithm converges under some mild conditions. The resultant estimator is shown to be asymptotically efficient in the sense of \cite{bickel1993efficient}, and it could be more powerful than the existing methods that do not incorporate covariates in some situations. Furthermore, our expectation-maximization algorithm can be naturally extended to handle missing child genotypes.

\section{Method}
\label{s:method}

\subsection{Model and notations}
Let $Y$ denote binary case-control status ($Y = 1$: case; $Y = 0$: control), ($g^m$, $g^c$) denote the target SNP genotypes for a mother-child pair, and $X$ denote a $p$-dimensional vector of environmental factors (covariates) collected from the mother. Let $A$ and $a$ denote the two alleles of the target SNP, where $a$ is the minor allele (frequency $\le0.5$). For the sake of simplicity, we assume that all SNPs follow an additive mode of inheritance by coding the genotypes $AA$, $Aa$, and $aa$ as $0$, $1$, and $2$, respectively. Let $g^c_m$ and $g^c_p$ denote children alleles at the target SNP inherited from the mother and father, respectively, coded as $0$ and $1$ for $A$ and $a$, respectively. The parental origin information can be unambiguously inferred if and only if at least one of $g^m$ and $g^c$ is not 1. In the situation where the parental origin information cannot be unambiguously inferred using the target SNP genotypes, one can use the genotypes of those SNPs tightly linked with the target SNP to improve inference accuracy. Such SNPs are called adjacent SNPs hereafter. Let $G^m$ and $G^c$ denote the genotype vectors at the target SNP and adjacent SNPs for the mother and child, respectively.

Let $n$ denote the total number of families. For $u=1,\ldots,n$, let the observed data for the $u$th mother-child pair be denoted by $(Y_u, X_u, G_u^m, G_u^c)$. We assume the following penetrance function: 
\begin{equation} \label{equ:penetrance}
\pr(Y=1 \mid g^{m}, g_{m}^{c}, g_{p}^{c}, X)= \mbox{expit}\{H(g^{m}, g_{m}^{c}, g_{p}^{c}, X ; \bm{\beta})\}, 
\end{equation}
where $\mbox{expit}(\cdot) = e^\cdot/(1 + e^\cdot)$ is the expit function, $H(g^{m}, g_{m}^{c}, g_{p}^{c}$, $X$ ; $\bm{\beta})$ can be a linear combination function of $g^m$, $g^c$, $g^c_m-g^c_p$, $X$, and their products, and $\bm{\beta}$ is a $p$-vector of associated parameters. For example, a penetrance function involving main genetic effects, parental origin effects, and gene-environment interaction effects is
\begin{align}
H\left(g^{m}, g_{m}^{c}, g_{p}^{c}, X ; \bm{\beta}\right)= &\beta_0 + \beta_{g^{m}} g^{m} + \beta_{g^{c}} g^{c} + \beta_{i m}\left(g_{m}^{c}-g_{p}^{c}\right) + \beta_{X}^{\T} X \notag\\
&+ \beta_{g^m X}^{\T} g^{m} X + \beta_{g^c X}^{\T} g^{c} X,\label{equ:penetrance1}
\end{align}
where $a^\T$ stands for the transpose of vector $a$.
We are interested in detecting the parental origin effect $\beta_{im}$. 

\subsection{A retrospective likelihood function for observed data}

Let $(Y_u,X_u,G^m_u,G^c_u,g^m_u,g^c_u,g^c_{mu},g^c_{pu})$, $u=1,\ldots,n$, be independent copies of the random vector $(Y,X,G^m,G^c,g^m,g^c,g^c_{m},g^c_{p})$.
Throughout this paper, we assume that the population prevalence $f$ is known \textit{a priori}. Then, under the case-control design, the retrospective likelihood function can be written as
\[\prod_{u=1}^{n} \pr\left(G_{u}^{m}, G_{u}^{c}, X_{u}\mid Y_{u}\right)\]
under the constraint $\pr(Y=1) = f$.
Recall that $g^m_u$, the maternal genotype at the target SNP, is a component of the maternal genotype vector $G^m_u$ so that $\pr(G^m_u,g^m_u)=\pr(G^m_u)$. Then the retrospective likelihood function is equivalent to 
\begin{equation} \label{equ:retro}
\prod_{u=1}^{n} \pr\left(Y_{u},G_{u}^{m}, G_{u}^{c}, X_{u}  \right)
=\prod_{u=1}^{n}\pr(Y_u\mid G_u^m,G_u^c,X_u)\pr(G_u^m,G_u^c\mid g_u^m,X_u)\pr(g_u^m,X_u) 
\end{equation}
under the constraint $\pr(Y=1) = f$. 

In our method, we leave the joint distribution of $(g^m_u,X_u)$ to be totally unspecified. That is, the empirical likelihood approach \citep{owen2001empirical} is adopted by introducing probability parameters $\pi_u=\pr(g^m_u,X_u)$ subject to the constraint $\sum_{u=1}^{n}\pi_u = 1$. Denote $\bm{\pi}=(\pi_1,\dots,\pi_n)^\T$.  We assume that $(G^m,G^c)$ and $X$ are conditionally independent given $g^m$, which holds true if the considered SNPs are associated with $X$ only through the target SNP \citep{chen2012semiparametric, zhang2020efficient}. Under this conditional independence assumption, we have 
\begin{equation} \label{equ:geno}
\pr(G^m,G^c\mid g^m,X)=\frac{\pr(G^m,G^c)}{\pr(g^m)}.
\end{equation}
Note that $\pr(Y\mid G^m,G^c,X)$ depends on $(G^m,G^c,X)$ only through the penetrance function $\pr(Y=1\mid g^m,g^c_m,g^c_p,X)$. However, $(g^c_m,g^c_p)$ cannot be inferred only using the genotypes of mother-child pair at the target SNP when both of the genotypes are heterozygous. Instead, we can apply the Law of Total Probability to calculate $\pr(Y\mid G^m,G^c,X)$ by going through all possible haplotype combinations compatible with the observed genotypes. Assume the joint genotypes are availabe for the target SNP and its $K-1$ adjacent SNPs, and any two of the $K$ SNPs are close enough so that the recombination probability between them is nearly 0. Let $\{h_1,\dots,h_S\}$ denote all possible haplotypes, and let $\bm{\mu} = (\mu_1,\dots,\mu_S)^\T$ denote the corresponding haplotype frequencies with $\sum_{s=1}^{S} \mu_s = 1$. Let $h_i^m h_j^m$ denote an unordered diplotype (haplotype pair) for the mother, and $(h_w^c,h_l^c)$ denote an ordered diplotype for the child,
where $h_w^c$ is the haplotype inherited from the mother so that $w=i$ or $w=j$, and $h_i^m$, $h_j^m$, $h_w^c$, and $h_l^c$ belong to $\{h_1,\dots,h_S\}$. The retrospective likelihood function \eqref{equ:retro} can be rewritten as
\begin{equation} \label{equ:haplo}
\prod_{u=1}^{n}\sum_
{\substack{h_{iu}^{m}h_{ju}^{m}\in{H}_u^m\\(h_{wu}^c,h_{lu}^c)\in{H}_u^c}}
\pr(Y_u\mid h_{iu}^{m}h_{ju}^{m},h_{wu}^{c},h_{lu}^{c},X_u)\pr(h_{iu}^{m}h_{ju}^{m},h_{wu}^{c},h_{lu}^{c}\mid g_u^m,X_u)\pr(g_u^m,X_u).
\end{equation}
Here, $H_u^{m}=\{h_{i u}^{m} h_{j u}^{m}\mid i,j=1,\ldots,S\}$ is the set of unordered diplotypes compatible with the joint maternal genotype $G_{u}^{m}$ for the $u$th family, and $H_u^{c}=\{(h_{w u}^{c}, h_{l u}^{c})\mid {w,l}=1,\ldots,S\}$ 
is the set of ordered diplotypes compatible with the joint maternal and child genotypes $(G_{u}^{m},G_{u}^{c})$ for the $u$th family. 

Let $\bmTheta = (\bm{\beta}^\T,\bm{\mu}^\T)^\T$ denote the vector of all unknown model parameters. Similar to \eqref{equ:geno}, we have 
\[\pr(h_i^m h_j^m,h_w^c,h_l^c\mid g^m,X)=\frac{\pr(h_i^m h_j^m,h_w^c,h_l^c)}{\pr(g^m)}, \]
so that \eqref{equ:haplo} can be written as 
\begin{equation} \label{equ:semi}
L(\bmTheta,\bm{\pi}) = \prod_{u=1}^{n}\sum_
{\substack{h_{iu}^{m}h_{ju}^{m}\in{H}_u^m\\(h_{wu}^c,h_{lu}^c)\in{H}_u^c}}
\pr(Y_u\mid h_{iu}^{m}h_{ju}^{m},h_{wu}^{c},h_{lu}^{c},X_u)\frac{\pr(h_{iu}^m h_{ju}^m,h_{wu}^c,h_{lu}^c)}{\pr(g_u^m)}\pi_u.
\end{equation}
We assume that the outcome depends on the haplotypes and diplotypes only through the genotypes at the target SNP, i.e., $\pr(Y_u\mid h_{iu}^{m}h_{ju}^{m},h_{wu}^{c},h_{lu}^{c},X_u) = \pr(Y_u\mid g_u^m,g_{mu}^c,g_{pu}^c,X_u)$, which is a function of $\bm{\beta}$. The probability of any possible haplotype combination can be obtained under the assumptions of Mendelian law and Hardy-Weinberg equilibrium for haplotypes, i.e., $\pr(h_{iu}^m h_{ju}^m,h_{wu}^c,h_{lu}^c) = \mu_i \mu_j \mu_l$, which is a function of $\bm{\mu}$. The target SNP minor allele frequency $\theta$ is a function of $\bm{\mu}$, so $\pr(g_u^m)$ is also a function of $\bm{\mu}$. In the next subsection, we will derive a modified profile likelihood to avoid estimating the high-dimensional nuisance parameters $\bm{\pi}$.

\subsection{An efficient estimator based on a modified profile log-likelihood}
\label{sec:modified}
Unknown parameters can be estimated by maximizing the likelihood function \eqref{equ:semi} under constraints $\pr(Y=1)=f$ and $\sum_{u=1}^{n}\pi_u = 1$, or equivalently, maximizing the log-likelihood function
\[\log L(\bmTheta,\bm{\pi}) = l_1(\bmTheta) + \sum_{u=1}^{n}\log \pi_u \]
under the same constraints, where $l_1(\bmTheta)$ is defined as
\begin{equation}\label{l1}
l_1(\bmTheta) = \sum_{u=1}^{n}\log\sum_{\substack{h_{iu}^{m}h_{ju}^{m}\in{H}_u^m\\(h_{wu}^c,h_{lu}^c)\in{H}_u^c}} \left[\pr(Y_u\mid h_{iu}^{m}h_{ju}^{m},h_{wu}^{c},h_{lu}^{c},X_u)\frac{\pr(h_{iu}^{m}h_{ju}^{m},h_{wu}^{c},h_{lu}^{c})}{\pr(g_u^m)}\right]. 
\end{equation}
The first constraint can be rewritten as 
\begin{equation}
\sum_{u=1}^{n} L_u(\bmTheta) \pi_{u}=f,
\end{equation}
where
\begin{equation}\label{Hu}
L_u(\bmTheta) := \pr(Y=1\mid g_u^m,X_u) = \sum_{(g_{mu}^c,g_{pu}^c)} \pr(Y=1\mid g^m_u,g_{mu}^c,g_{pu}^c,X_u)\pr(g_{pu}^c)\pr(g_{mu}^c\mid g^m_u).
\end{equation}
By virtue of the Lagrange multiplier method, the above constrained optimization problem can be shown to be equivalent to maximizing the profile log-likelihood
\begin{equation} \label{equ:profile}
l_p(\bmTheta) = l_0(\bmTheta,\lambda_{\bmTheta} ) = l_1(\bmTheta) - \sum_{u=1}^{n} \log \left[n\left\{1+\lambda_{\bmTheta}\left(L_u(\bmTheta)-f\right)\right\}\right],
\end{equation}
where $\lambda_{\bmTheta}$ is the solution to the equation 
\begin{equation}
\sum_{u=1}^{n} \frac{L_u(\bmTheta)-f}{1+\lambda\left(L_u(\bmTheta)-f\right)}=0 \end{equation}
with respect to $\lambda$. The derivation of \eqref{equ:profile} is postponed to Appendix A (Supplementary Material).

We can solve the score equations
\begin{equation}
\frac{\partial l_0(\bmTheta, \lambda)}{\partial \lambda}=0 \ \mbox{and}\  \frac{\partial l_0(\bmTheta, \lambda)}{\partial \bmTheta}=0
\end{equation}
to obtain the maximizer of $l_p(\bmTheta)$ defined in \eqref{equ:profile}. However, $l_0(\bmTheta, \lambda)$ is not a log-likelihood function, and the solution to the above score equations is a saddle point, as can be shown similarly to \cite{zhang2018efficient}. As in \cite{zhang2018efficient,zhang2020efficient}, to resolve this numerical problem, we propose to modify the profile log-likelihood function \eqref{equ:profile} by replacing the multiplier $\lambda_{\bmTheta}$ with its limiting value
\begin{equation}
\lambda_0 =\frac{n_1}{nf} - \frac{n_0}{n(1-f)}.    
\end{equation}
Accordingly, the maximization problem is simplified to maximizing the following modified profile log-likelihood function:
\begin{equation}\label{equ:modified}
l_{\operatorname{mp}}(\bmTheta) =l_1(\bmTheta) - l_2(\bmTheta,\lambda_0),
\end{equation}
where 
\begin{equation}\label{equ:l2}
l_2(\bmTheta,\lambda_0)=\sum_{u=1}^{n} \log \left[n\left\{1+\lambda_0\left(L_u(\bmTheta)-f\right)\right\}\right].
\end{equation}
An estimator $\hat\bmTheta_{\operatorname{mp}}$ can be obtained by solving the corresponding score equation
\begin{equation}\label{equ:scoreequation}
    \frac{\partial l_{\operatorname{mp}}(\bmTheta)}{\partial \bmTheta} = 0.
\end{equation}
In what follows, we establish some large sample properties of $\hat\bmTheta$. Let $\bmTheta_{0}$ denote the true value of $\bmTheta$. Denote $Z = (G^m,G^c,X)$, $W = (g^m,X)$, $p_{Z}(Z ; \bmTheta)=\pr_{\bmTheta}(Y=1\mid Z)$. We need the following regularity conditions:
\begin{description}
\item[] (C1) The parameter space is $\{(\bm{\beta}, \bm{\mu})\mid \bm{\beta} \in \mathbb{R}^p; \bm{\mu}=(\mu_1, \ldots, \mu_S)^\T \text{ with }\mu_1, \ldots, \mu_S > 0$ and $\sum_{s=1}^{S} \mu_s = 1\}$, and the true parameter vector $\bmTheta_0$ is an interior point of the parameter space. 
\item[] (C2) The expectations
$$E\bigg[\bigg|\frac\partial{\partial \bmTheta} p_{Z}(Z ; \bmTheta)\bigg|_{\bmTheta=\bmTheta_{0}}\bigg|\bigg]$$ 
and
$$E\bigg[\bigg|\frac{\partial}{\partial \bmTheta} \log\pr_{\bmTheta}(G^m,G^c\mid W)\bigg|_{\bmTheta=\bmTheta_{0}}\bigg|\bigg]$$
exist.
\item[] (C3) The sample size ratio $n_1/n_0$ is fixed as long as the total sample size $n=n_0+n_1$ varies. That is, there exists a constant $a_0\in(0,1)$ independent of $n$, such that $n_0 = a_0 n$ and $n_1 = a_1 n$, where $a_1=1-a_0$. 

\item[] (C4) The expectation $E[|l_{\operatorname{mp}}(\bmTheta)|]$ exists for any $\bmTheta$ in some neighborhood of $\bmTheta_0$.

\item[] (C5) Minus the expected Hessian matrix
$$-E\bigg[\bigg|\frac{\partial^{2}l_{\operatorname{mp}}(\bmTheta)}{\partial \bmTheta \partial \bmTheta^\T} \bigg|\bigg]\bigg|_{\bmTheta=\bmTheta_{0}} $$ 
is positive definite.

\item[] (C6) For any $\bmTheta$ in some neighborhood of $\bmTheta_0$ and the $i$th element $\Theta_i$ and $j$th element $\Theta_j$ of $\bmTheta$, there exists a non-negative random variable $C$ with finite expectation, such that 
$$\bigg|\frac{\partial^{2}l_{\operatorname{mp}}(\bmTheta)}{\partial \Theta_i \partial \Theta_j}  - \frac{\partial^{2} l_{\operatorname{mp}}(\bmTheta)}{\partial \Theta_i \partial \Theta_j}\bigg|_{\bmTheta=\bmTheta_{0}}\bigg| \le C\|\bmTheta - \bmTheta_0\|.$$

\item[] (C7) The so called ``Fisher information matrix''
$$\operatorname{cov}\bigg(\frac{\partial l_{\operatorname{mp}}(\bmTheta)}{\partial \bmTheta}\bigg)\bigg|_{\bmTheta=\bmTheta_{0}}$$
exists.

\item[] (C8) The following technical condition holds: 
$$E\bigg[\frac{\partial^{2} l_0(\bmTheta, \lambda)}{\partial \bmTheta \partial \lambda^{\T}}\bigg]\bigg|_{\bmTheta=\bmTheta_{0},\lambda=\lambda_{0}}=0.$$
\end{description}

In what follows, we first show in Lemma \ref{lemma1} that the function $l_{\operatorname{mp}}(\bmTheta)$ shares a property with the conventional likelihood score, then present our main asymptotic results in Theorem \ref{theorem1}. Refer to Appendices B and C (Supplementary Material) for proofs of Lemma \ref{lemma1} and Theorem \ref{theorem1}.

\begin{lemma}
\label{lemma1}
Under the regularity condition (C2), we have
$$
\left.E\left[\frac{\partial l_{\operatorname{mp}}(\bmTheta)}{\partial \bmTheta}\right]\right|_{\bmTheta=\bmTheta_{0}}=0.
$$
\end{lemma}

\begin{theorem}\label{theorem1}
As the total sample size $n$ goes to infinity, we have the following asymptotic properties:
\begin{description}
\item [](i) Under condition (C1), with probability tending to 1, $l_{\operatorname{mp}}(\bmTheta)$ takes the value of negative infinity on the boundary of the parameter space, and 
there exists a solution to \eqref{equ:scoreequation}, denoted by $\hat{\bmTheta}_{\operatorname{mp}}$, in the interior of the parameter space.
\item [](ii) Under conditions (C1)-(C5),  $\hat{\bmTheta}_{\operatorname{mp}}$  converges in probability to $\bmTheta_0$.
\item [](iii) Under conditions (C1)-(C7),
$$n^{1/2}(\hat{\bmTheta}_{\operatorname{mp}}-\bmTheta_0)\rightarrow N\left(\bf{0},A_{\operatorname{mp}}^{-1}(\bmTheta_{0}) \Sigma_{\operatorname{mp}}\left(\bmTheta_{0}\right) A_{\operatorname{mp}}^{-1}\left(\bmTheta_{0}\right)\right)\text{ in distribution},$$
where 
$$A_{\operatorname{mp}}(\bmTheta_{0})=\frac1n E\bigg[\frac{\partial^{2} l_{\operatorname{mp}}(\bmTheta)}{\partial \bmTheta \partial \bmTheta^\T}\bigg]\bigg|_{\bmTheta=\bmTheta_{0}}$$
and  
$$\Sigma_{\operatorname{mp}}(\bmTheta_{0})=\frac1n \operatorname{cov}\bigg( \frac{{\partial l_{\operatorname{mp}}(\bmTheta)}}{\partial \bmTheta}\bigg)\bigg|_{\bmTheta=\bmTheta_{0}}.$$

\item [] (iv) Under conditions (C1)-(C8), $\hat{\bmTheta}_{\operatorname{mp}}$ is asymptotic equivalent to the conventional maximum likelihood estimator, which means that $\hat{\bmTheta}_{\operatorname{mp}}$ is asymptotically efficient in the sense of \cite{bickel1993efficient}.
\end{description}
\end{theorem}

Note that $A_{\operatorname{mp}}(\bmTheta_{0})$  can be consistently estimated by 
$$\hat{A}_{\operatorname{mp}}(\hat{\bmTheta}_{\operatorname{mp}}) := \frac1n\frac{\partial^{2} l_{\mathrm{\operatorname{mp}}}(\bmTheta)}{\partial \bmTheta \partial \bmTheta^{\T}}\Big|_{\bmTheta=\hat{\bmTheta}_{\operatorname{mp}}},$$
and $\Sigma_{\operatorname{mp}}(\bmTheta_{0})$ can be consistently estimated by a weighted average of two sample variance-covariance matrices corresponding to the subgroups of cases and controls. Consequently, $n^{1/2}(\hat{\bmTheta}_{\operatorname{mp}}-\bmTheta_0)$ is approximately multivariate normal with zero expectation and variance-covariance matrix $\hat{A}_{\operatorname{mp}}^{-1}(\hat{\bmTheta}_{\operatorname{mp}}) \hat{\Sigma}_{\operatorname{mp}}(\hat{\bmTheta}_{\operatorname{mp}})\hat{A}_{\operatorname{mp}}^{-1}(\hat{\bmTheta}_{\operatorname{mp}})$.
Confidence intervals and significance tests of $\bmTheta$ can be constructed based on this normality approximation. 

\begin{remark}
Condition (C2) is a sufficient condition for the existence of $E[\partial l_{\operatorname{mp}}(\bmTheta_0) / \partial \bmTheta]$, which is similar to that for the conventional likelihood function. Condition (C3) is commonly used to establish large sample properties for case-control studies \citep{bickel1993efficient}. Conditions (C4)-(C7) ensure the consistency and asymptotic normality of $\hat{\bmTheta}_{\operatorname{mp}}$, which are also similar to those for the conventional likelihood function.
\end{remark}

\begin{remark}
Condition (C8) is used in the proof of the asymptotic equivalence of $\hat{\bmTheta}_{\operatorname{mp}}$ and the asymptotically efficient estimator (i.e., the maximizer of the profile likelihood function \eqref{equ:profile}), so that $\hat{\bmTheta}_{\operatorname{mp}}$ is also asymptotically efficient. We demonstrated that this condition is nearly satisfied in some simulation situations, refer to Appendix D and Table S1 (Supplementary Material) for details. 
\end{remark}

\subsection{An expectation-maximization algorithm for the modified profile log-likelihood}\label{EMalgorithm}

Note that the profile log-likelihood \eqref{equ:modified} has a complex form, whose maximization is computationally intensive. By treating haplotypes as missing data, the maximization of \eqref{equ:modified} can be effectively simplified by applying an expectation-maximization algorithm. Specifically, the expectation step is to calculate the conditional expectation of the modified profile log-likelihood function given the observed data (genotype data, disease status, and covariates) under initial parameters or those estimated in the previous step, and the maximization step is to maximize the resultant conditional expectation with respect to unknown parameters. This expectation-maximization algorithm is detailed as follows.

Let $\bmTheta^{(r)}$ denote the estimator of $\bmTheta$ obtained at the $r$th iteration of the expectation-maximization algorithm, and let $\mathcal{D}$ denote all observed data, $\mathcal{H}$ denote the missing haplotype data. Then, the modified profile log-likelihood for the complete data ($\mathcal{D}$, $\mathcal{H}$) is
\begin{equation}\label{equ:lmpc}
l_{\operatorname{mp}}(\bmTheta;\mathcal{D},\mathcal{H})=l_1(\bmTheta;\mathcal{D},\mathcal{H})-l_2(\bmTheta,\lambda_0),
\end{equation}
where
\begin{equation} \label{equ:EMcom}
	l_{1}(\bmTheta;\mathcal{D},\mathcal{H}=h) = \sum_{u=1}^{n} \log \pr(Y_{u}, G_{u}^{c}, G_{u}^{m},h_u\mid g_u^m, X_{u}) = \sum_{u=1}^{n}\log\bigg[\pr(Y_u\mid X_u,h_u)\frac{\pr(h_u)}{\pr(g_u^m)}\bigg],
\end{equation}
and $h_u$ is the haplotype vector for the $u$th family, which is compatible with the observed joint mother-child genotypes under some regularity conditions such as tight linkage, random mating, and Mendelian inheritance.
Noting that $L_u(\bmTheta)$ does not depend on the missing haplotype data, we have $E_{\bmTheta^{(r)}}[l_2(\bmTheta,\lambda_0)\mid \mathcal{D}]=l_2(\bmTheta,\lambda_0)$. Consequently, in the $(r+1)$th iteration of the expectation-maximization algorithm, the expectation step is to compute
\begin{equation} \label{equ:estep}
E_{\bmTheta^{(r)}}[l_{\operatorname{mp}}(\bmTheta;\mathcal{D},\mathcal{H})\mid \mathcal{D}] = E_{\bmTheta^{(r)}}[l_{1}(\bmTheta;\mathcal{D},\mathcal{H})\mid\mathcal{D}] - l_2(\bmTheta,\lambda_0),
\end{equation}
where 
$$E_{\bmTheta^{(r)}}[l_{1}(\bmTheta;\mathcal{D},\mathcal{H})\mid\mathcal{D}]=\sum_{u=1}^n \sum_{h_u} pr_{\bmTheta^{(r)}}(h_u\mid Y_u,X_u,G_u^m,G_u^c)\log\bigg[\pr(Y_u\mid X_u,h_u)\frac{\pr(h_u)}{\pr(g_u^m)}\bigg].$$ 
The maximization step in the $(r+1)$th iteration of the expectation-maximization algorithm is to maximize \eqref{equ:estep} with respect to $\bmTheta$ using any non-linear optimization algorithm.

The above expectation-maximization algorithm is different from the conventional expectation-maximization algorithm since our objective function is a modified profile log-likelihood function instead of a log-likelihood function. Nevertheless, we still have the following convergence result for our expectation-maximization algorithm (refer to Appendix E in Supplementary Material for a proof):
\begin{theorem} \label{theorem2}
If $l_{mp}(\bmTheta_1) \neq l_{mp}(\bmTheta_2)$ for any two stationary points  $\bmTheta_1$ and $\bmTheta_2$ of the modified profile log-likelihood, then $\bmTheta^{(r)}$ defined below \eqref{equ:l2} converges to a stationary point of the modified profile log-likelihood as $r\to\infty$.
\end{theorem}

A starting value $\bmTheta^{(0)}$ is required for the proposed expectation-maximization algorithm. In practice, we propose to use a prospective logistic regression to be introduced in Section~\ref{s:simu} to obtain a reasonably good starting value $\bmTheta^{(0)}$, which can speed up the convergence of the expectation-maximization algorithm. The maximization step involves unknown parameters $\bmTheta=(\bm{\beta},\bm{\mu})$. To simplify the maximization problem, we propose to update only a part of parameters ($\bm{\beta}$ or $\bm{\mu}$) at a time \citep{meng1993maximum}.

Our proposed method can be naturally extended to handle missing child genotype data for a subset of families by slightly modifying \eqref{equ:modified}. That is, we allow the haplotype combinatorial sets $H_u^m$ and $H_u^c$ in \eqref{l1} to be compatible with the observed maternal genotype data only.

\subsection{Implementation of the proposed method}

We outline our inference procedure as follows.

\begin{enumerate}
\item [(1)] Specify the prevalence $f$. The prevalence can be either estimated from extra cohort studies or extracted from public databases. As demonstrated in our simulation study in Section~\ref{s:simu}, misspecifying $f$ only affects the estimation of intercept parameter $\beta_0$ and does not affect the other parameters of our interest.

\item [(2)] Specify possible haplotypes associated with the target SNP and available adjacent SNPs, which can be obtained by any existing method such as the one implemented in the function \texttt{haplo.em} in the \texttt{R} package \texttt{haplo.stats}. 

\item [(3)] Specify an initial value of haplotype frequency vector $\bm{\mu}$. The outputs of the \texttt{R} function \texttt{haplo.em} include an estimate of $\bm{\mu}$, which can serve as an initial value. 

\item [(4)] Specify an initial value of regression parameter vector $\bm{\beta}$. A prospective logistic
regression method to be introduced in Section~\ref{s:simu} can be applied to produce an estimate of $\beta$, which can serve as an initial value of $\bm{\beta}$. 

\item [(5)] Apply the expectation-maximization algorithm described in Section \ref{EMalgorithm} to estimate both $\bm{\mu}$ and $\bm{\beta}$, with the initial values being specified in steps (3) and (4), and estimate the variance matrix of $\hat{\bm{\beta}}$ as described in Section~\ref{sec:modified}. 

\item [(6)] Obtain confidence intervals (z-intervals) and p-values of significance test (z-tests) for $\bm{\beta}$.
\end{enumerate} 

For convenience, we refer to our robust method based on the modified profile log-likelihood function as ROB-HAP. Here, ROB is the shorthand for robust, emphasizing that our method is robust in the sense that it does not impose any constraint on the relationship between the maternal genotype and the covariates; HAP is the shorthand for haplotype, emphasizing that our method uses haplotypes to determine the parental origins.

Our proposed method ROB-HAP has been implemented in an \texttt{R} function \texttt{assessPOE}, which is available at \url{https://github.com/yatian20/APOE}. The inputs of \texttt{assessPOE} include the case-control status $Y$, maternal covariates $X$, genotypes of mother-child pairs $(G^m,G^c)$, prevalence $f$, and target SNP ID. The outputs of \texttt{assessPOE} include point estimates, confidence intervals, and significance test p-values for $\bm{\beta}$. In our preliminary numerical study, we found that it was quite computationally intensive to run our expectation-maximization algorithm when many SNPs were involved. To speed up the computation, users can adopt the \texttt{R} package \texttt{LDheatmap} \citep{shin2006ldheatmap} to select a limited number of SNPs with strong linkage disequilibrium among the candidate SNPs.

\section{Simulation Studies}\label{s:simu}

\subsection{Data generation} \label{GENdata}

Simulation studies were conducted to evaluate the performance of our proposed method ROB-HAP. The general data generation process is described in this subsection, then specific parameter settings are given in the subsequent subsections. We generated genotype data for five SNPs using seven identified haplotypes, which were extracted from published haplotype data in the genomic region GPX1. The configurations and frequencies of the haplotypes can be found in Table 1 of \cite{chen2004haplotype}. Diplotypes were then generated for $N = 10^7$ mother-child pairs under Hardy-Weinberg equilibrium. Specifically, three haplotypes $h_i^m$, $h_j^m$, $h_l^c$ were randomly sampled for each mother-child pair according to the distribution given in Table S2, then $h_w^c$ was generated according to the Mendelian law. Maternal genotype $g^m$ at the target SNP was extracted from the maternal diplotype. Environmental risk factor $X$ was generated through the following linear model:
\begin{equation} \label{equ:linear model}
X = \eta\{g^m - E(g^m)\} + e,
\end{equation}
where $E(g^m)$ was the expectation of $g^m$, the random error $e$ was independent of $g^m$ and followed the standard normal distribution, and $\eta$ was a regression parameter. The intercept $\beta_0$ was determined by the above distributions, the prevalence of phenotype and log-odds ratio (log-OR) parameters. Unless otherwise specified, we fixed $\eta = \log(3.0)$ in model \eqref{equ:linear model} and $\beta_{g^m} = \log(1.8)$, $\beta_{g^c} = \log(1.5)$, $\beta_{im} = \log(1.5)$, $\beta_{X} = \log(1.8)$, and $\beta_{g^mX}=\beta_{g^cX}=0$ in model \eqref{equ:penetrance1}. Hardy-Weinberg equilibrium was adopted in generating diplotypes unless otherwise specified. The phenotypes $Y$'s were generated according to the penetrance model \eqref{equ:penetrance} together with \eqref{equ:penetrance1}. Data were randomly sampled for $n_0 = 200$ control families and $n_1 = 200$ case families with $Y=0$ and $Y=1$, respectively.

\subsection{Considered methods}

For the comparison purpose, we considered several alternative methods to ROB-HAP. The first alternative method is similar to ROB-HAP but it directly uses the parental origin information that could not be completely inferred in practice, which is referred to as ROB-COM (COM is the shorthand for complete). The second alternative method is a special version of ROB-HAP that only uses the genotypes from the target SNP instead of all available SNPs to infer the parental origins, which is referred to as ROB-SNP. The third alternative method is a simplified version of ROB-HAP by further incorporating independence assumption between maternal genotype and covariates, which is referred to as IND-HAP (IND is the shorthand for independent). In terms of inference efficiency, we can intuitively rank the four methods mentioned so far: ROB-COM $>$ IND-HAP $>$ ROB-HAP  $>$  ROB-SNP. However, one should be aware that ROB-COM is not applicable in practice if the parental origins cannot be unambiguously inferred for some families and IND-HAP could produce biased results if the independence assumption is violated. We also considered the prospective logistic regression, which is referred to as LOGIT-HAP (the fourth alternative method). In LOGIT-HAP, first, the joint diplotypes of mother-child pairs are inferred using the function \texttt{haplo.em} in the \texttt{R} package \texttt{haplo.stats} with the adjacent SNP genotypes, then, the parental origins are determined by the joint diplotypes and only those families with uniquely inferred parental origins are kept. Since LOGIT-HAP is based on a prospective likelihood, it is not efficient by ignoring the information on Hardy-Weinberg equilibrium, Mendelian law between maternal genotype and child genotype, and the conditional independence between covariates and child genotype given maternal genotype. Finally, we considered two existing methods in the literature, i.e., the fifth alternative method P-HAP \citep{lin2013multi} and the sixth alternative method EMIM \citep{ainsworth2011investigation}. Both P-HAP and EMIM can evaluate parental origin effects using case-control mother-child pair data but neither of them can incorporate covariate information. The \texttt{R} package HAPLIN can also be used to detect parental origin effects, which is based on a log-linear model \citep{gjessing2006case}. It has been shown that EMIM and Haplin have comparable powers in detecting genetic effects through extensive simulation studies \citep{gjerdevik2019haplin}, and EMIM is more often adopted due to its computational advantage. We therefore do not show the simulation results for HAPLIN in the next sections.

\subsection{Comparison of prospective and retrospective likelihoods}\label{simu1}

We conducted a simulation study to demonstrate the advantage of the retrospective likelihood (ROB-HAP) over the prospective likelihood (LOGIT-HAP). The third of the five SNPs was set as the target SNP unless specified otherwise. Estimation results based on 2,500 simulated datasets are summarized in the upper panel of Table \ref{tab1}.

\begin{table}	
    \caption{
		Estimation results with $g^m$ and $X$ being correlated ($\eta = \log 3$; $f = 0.01$)
		}\label{tab1}
    \centering
	\begin{threeparttable}
        \begin{tabular*}{\hsize}{@{}@{\extracolsep{\fill}}ccccccccccc@{}} 
			\hline
			&       & \multicolumn{4}{c}{ROB-HAP$^a$}    &  & \multicolumn{4}{c}{LOGIT-HAP$^b$}  \\ \cline{3-6} \cline{8-11} 
			Log-OR        & True$^c$  & Bias$^d$   & SE$^e$    & SEE$^f$   & CP$^g$    &  & Bias$^d$   & SE$^e$    & SEE$^f$   & CP$^g$    \\ \hline
			$\beta_{g^m}$ & 0.588 & 0.017  & 0.259 & 0.261 & 0.955 &  & 0.029  & 0.304 & 0.304 & 0.951 \\
			$\beta_{g^c}$ & 0.405 & -0.012 & 0.177 & 0.174 & 0.948 &  & 0.018  & 0.245 & 0.237 & 0.945 \\
			$\beta_{im}$  & 0.405 & -0.003 & 0.188 & 0.187 & 0.949 &  & -0.013 & 0.270 & 0.265 & 0.952 \\
			$\beta_X$     & 0.588 & 0.011  & 0.124 & 0.124 & 0.953 &  & 0.017  & 0.126 & 0.125 & 0.951 \\ 
		\hline
		&       & \multicolumn{4}{c}{ROB-SNP$^i$}    &  & \multicolumn{4}{c}{ROB-COM$^j$}   \\ 
		\cline{3-6} \cline{8-11} 
		Log-OR        & True$^c$  & Bias$^d$   & SE$^e$    & SEE$^f$   & CP$^g$    &  & Bias$^d$  & SE$^e$    & SEE$^f$   & CP$^g$    \\ \hline
		$\beta_{g^m}$ & 0.588 & -0.009 & 0.266 & 0.270 & 0.958 &  & 0.006 & 0.256 & 0.257 & 0.953 \\
		$\beta_{g^c}$ & 0.405 & 0.002  & 0.179 & 0.178 & 0.953 &  & 0.002 & 0.176 & 0.172 & 0.946 \\
		$\beta_{im}$  & 0.405 & 0.011  & 0.216 & 0.218 & 0.954 &  & 0.006 & 0.174 & 0.172 & 0.951 \\
		$\beta_X$     & 0.588 & 0.007  & 0.124 & 0.125 & 0.957 &  & 0.011 & 0.126 & 0.125 & 0.947 \\ 
		\hline
		\end{tabular*}
		\begin{tablenotes}
\item $^a$Our proposed method using haplotypes; $^b$the conventional logistic regression method using haplotypes; $^c$the true value of the log-OR parameter; $^d$difference between the mean estimate and true parameter value; $^e$empirical standard error; $^f$mean estimated standard error; $^g$empirical coverage probability of the 95$\%$ confidence intervals; $^i$a simplified verion of ROB-HAP only using the genotypes at the target SNP; $^j$an ideal method utilizing the parental origin information that could not be completely inferred in practice.
		\end{tablenotes} 
	\end{threeparttable}
\end{table}

\begin{figure}
    \centering 
    \includegraphics[width=1\textwidth]{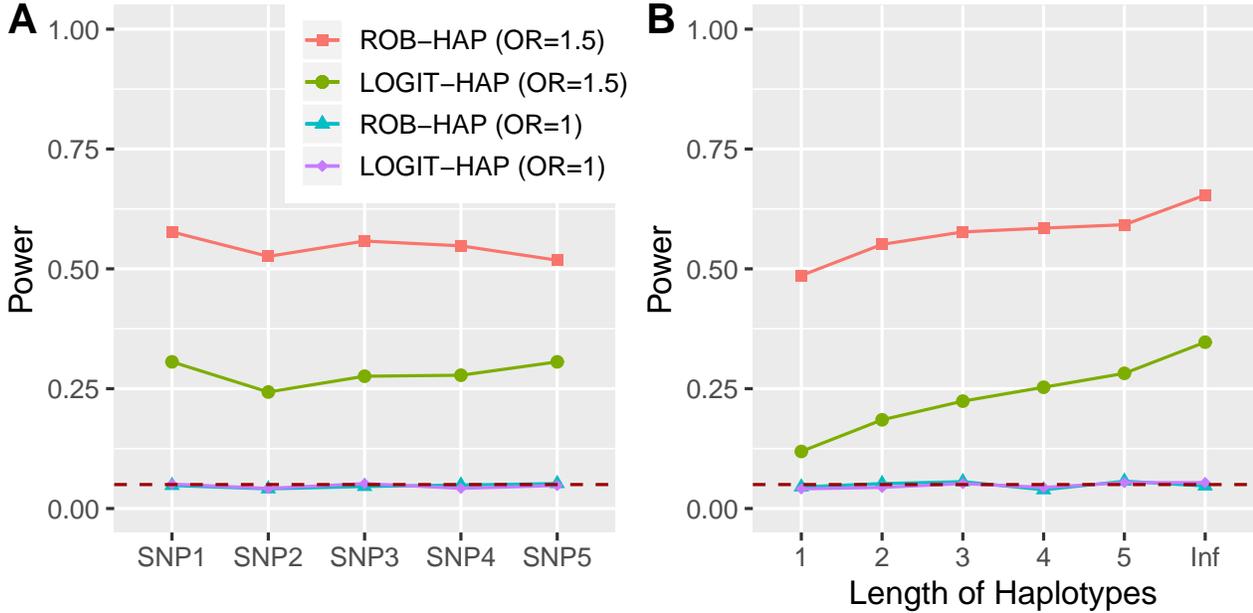} 
    \caption{Panel A: The type-I error rates and powers for testing parental origin effects with confounding covariate $X$ incorporated ($\eta=\log3$). Various target SNPs in gene GPX1 were considered. Data were generated under $\beta_{im} = 0$ (for assessing type-I error rates) or $\beta_{im} = \log1.5$ (for assessing powers). The other underlying parameters were $f = 0.01$, $\beta_{g^m} = \beta_{X} = \log1.8$, and $\beta_{g^c} = \log1.5$. Panel B: The type-I error rate and power trends against the length of haplotypes for testing parental origin effects. The target SNP was SNP1 in gene GPX1. Here Inf corresponds to ROB-COM with the true parent origins being used. Data were generated under $\beta_{im} = 0$ (for assessing the type-I error rates) or $\beta_{im} = \log1.5$ (for assessing the powers). The other underlying parameters were $f = 0.01$, $\eta = \log3$, $\beta_{g^m} = \beta_{X} = \log1.8$, and $\beta_{g^c} = \log1.5$.}
    \label{fig1}
\end{figure}

Evidently, both ROB-HAP and LOGIT-HAP were virtually unbiased, though the bias of LOGIT-HAP was slightly larger. The empirical standard errors (SEs) appeared to be close to the averages of estimated SEs (SEEs). ROB-HAP was uniformly more efficient than LOGIT-HAP in terms of SEs. For example, the SE of $\hat{\beta}_{im}$ was $30\%$ lower than that of LOGIT-HAP. Panel A of Fig. \ref{fig1} showcases the powers and type-I error rates for testing parental origin effects, where each of the five SNPs was treated in turn as the target SNP while the genotypes from other SNPs were used to infer the parental origins. Both LOGIT-HAP and ROB-HAP had perfect controls of type-I error rates. ROB-HAP were uniformly more powerful than LOGIT-HAP, with power gains ranging from 0.21 to 0.28.

\subsection{The impact of parental origins on the power of parental origin effect detection}\label{simu2}

Determining the parental origins is a key step in testing parental origin effects. We evaluated the power improvement of the proposed method ROB-HAP over ROB-SNP, where ROB-SNP only uses the target SNP to infer the parental origins. We also included ROB-COM for comparison, which uses the true parental origin information that could not be completely inferred in practice. ROB-SNP can be regarded as a special case of ROB-HAP with the number of adjacent SNPs being zero.
Results based on 2,500 simulated datasets are summarized in the lower panel of Table \ref{tab1}.

Both ROB-SNP and ROB-COM appeared to be unbiased. As expected, ROB-HAP was more efficient than ROB-SNP, especially for the parental origin effects. In fact, the SE of $\hat{\beta}_{im}$ by ROB-HAP was $13\%$ lower compared with ROB-SNP. Of course, ROB-COM was most efficient, with a further SE reduction of $6\%$. Efficiency gain of ROB-HAP over ROB-SNP could depend on the number of adjacent SNPs, minor allele frequencies, and so on. As shown in Panel B of Fig. \ref{fig1}, the power for testing parental origin effects steadily increased with the length of haplotypes. LOGIT-HAP had the same power trend, but it was uniformly inferior to ROB-HAP.

\subsection{Comparison with existing methods that do not incorporate covariates}\label{simu3}

There are quite a few methods in the literature for testing parental origin effects, but none of them can incorporate covariates in a general way. Among these methods, P-HAP \citep{lin2013multi} can use haplotype information to improve estimation/test efficiency for parental origin effects; EMIM \citep{ainsworth2011investigation} has been widely used in testing maternal, imprinting, and interaction effects. We conducted a simulation study to compare their performances with our proposed method ROB-HAP. Because P-HAP and EMIM cannot incorporate covariates, they could be invalid if the covariates are associated with both maternal genotypes and the phenotype of interest. We also considered a slightly modified version of ROB-HAP, which imposes the independence constraint between the covariates and maternal genotypes and is referred to as IND-HAP.

\begin{figure}
    \centering 
    \includegraphics[width=1\textwidth]{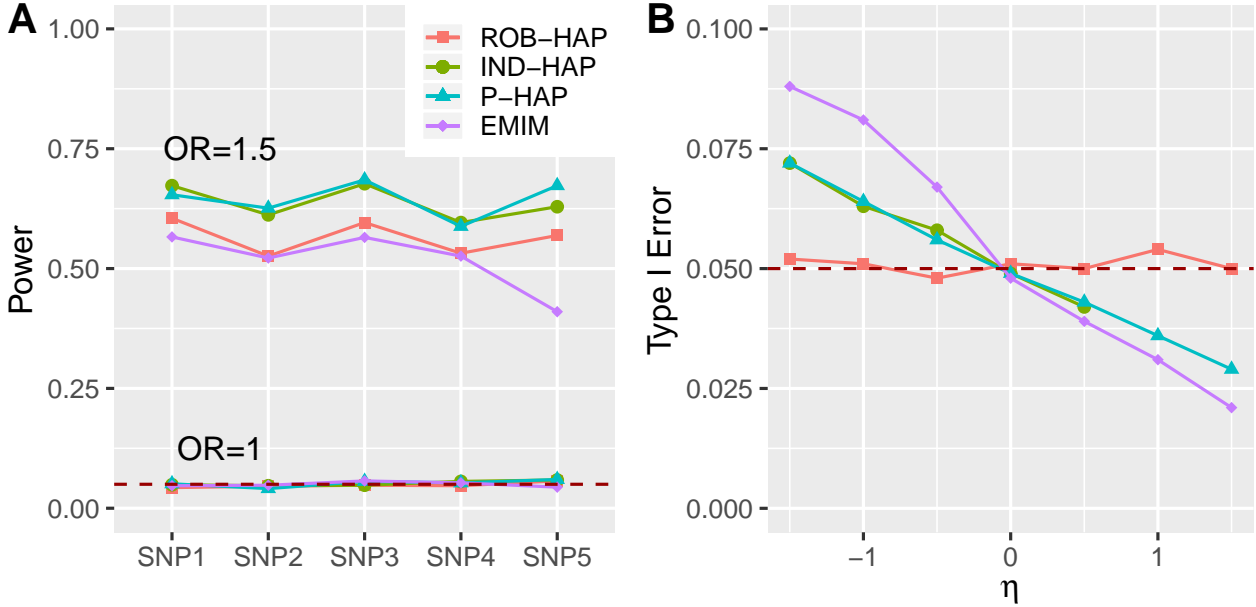} 
    \caption{Panel A: The type-I error rates (OR = 1) and powers (OR = 1.5) for testing parental origin effects with non-confounding covariate $X$ ($\eta=0$). The other underlying parameters were $f = 0.01$, $\beta_{g^m} = \beta_{X} = \log1.8$, and $\beta_{g^c} = \log1.5$. Panel B: Type-I error rates for testing parental origin effects in the presence of a confounding covariate $X$. The target SNP was SNP1 in gene GPX1. Data were generated under a non-linear model relating the covariate $X$ and maternal genotype $g^m$: $X = \eta \{(g^m)^2 - E(g^m)^2\} + e$, where $e$ was standard normal. The other underlying parameters were $f = 0.01$, $\beta_{im} = 0$, $\beta_{g^m} = \beta_{X} = \log1.8$, and $\beta_{g^c} = \log1.5$.}
    \label{fig2}
\end{figure}

Simulation data were generated as in Section \ref{simu1}, except that the covariate was independent of the maternal genotypes (i.e., $\eta=0$). The estimation results for IND-HAP and P-HAP are summarized in Table \ref{tab3}. Because EMIM was derived under a different model and the estimated parameters were different from those in model \eqref{equ:penetrance}, we only show the powers and type-I errors of EMIM with respect to the parental origin effects, but not its estimation results. Evidently, both IND-HAP and P-HAP were virtually unbiased, the estimated standard errors were close to the empirical ones, and the 95\% confidence intervals had coverage probabilities close to the nominal level. IND-HAP and P-HAP appeared to be comparable in terms of SEs. 
The type-I error rates and powers for testing parental origin effects are presented in Panel A of Fig. \ref{fig2} for IND-HAP, ROB-HAP, P-HAP, and EMIM.
The type-I error rates were well controlled by all the four methods. IND-HAP and P-HAP appeared to have comparable powers, EMIM was least powerful, and ROB-HAP was intermediate.

\begin{table}
	\centering
	\caption{Estimation results with $g^m$ and $X$ being independent ($\eta = 0$, $f = 0.01$)}\label{tab3}
	\begin{threeparttable}
        \begin{tabular*}{\hsize}{@{}@{\extracolsep{\fill}}ccccccccccc@{}} 
			\hline
			&       & \multicolumn{4}{c}{IND-HAP$^a$}    &  & \multicolumn{4}{c}{P-HAP$^b$}      \\ \cline{3-6} \cline{8-11} 
			Log-OR        & True$^c$  & Bias$^d$   & SE$^e$    & SEE$^f$   & CP$^g$    &  & Bias$^d$   & SE$^e$    & SEE$^f$   & CP$^g$    \\ \hline
			$\beta_{g^m}$ & 0.588 & -0.015 & 0.196 & 0.200 & 0.954 &  & 0.018  & 0.199 & 0.198 & 0.948 \\
			$\beta_{g^c}$ & 0.405 & 0.013  & 0.153 & 0.153 & 0.947 &  & 0.009  & 0.155 & 0.153 & 0.946 \\
			$\beta_{im}$  & 0.405 & -0.006 & 0.162 & 0.166 & 0.954 &  & -0.022 & 0.168 & 0.166 & 0.942 \\
			$\beta_X$     & 0.588 & 0.014  & 0.108 & 0.111 & 0.947 &  & - $^h$     & -     & -     & -     \\ \hline
    	\end{tabular*}
        \begin{tablenotes}
        \item $^a$A simplified version of ROB-HAP with independence assumption between $X$ and $g^m$; $^b$a parental origin effect testing method without incorporating covariates \citep{lin2013multi}; $^c$true value of the log-OR parameter; $^d$difference between the mean estimate and true parameter value; $^e$empirical standard error; $^f$mean estimated standard error; $^g$empirical coverage probability of the 95$\%$ confidence interval; $^h$covariates not incorporated in P-HAP. 
        \end{tablenotes} 
    \end{threeparttable}
\end{table}

When $g^m$ and $X$ was correlated ($\eta\not=0$), the three methods not incorporating $X$ or assuming $g^m$-$X$ independence could be either liberal or conservative, depending on $\eta<0$ or $\eta>0$ (Panel B of Fig. \ref{fig2}).

\subsection{Sensitivity analyses with model misspecification}

Some model assumptions are required for ROB-HAP, such as Hardy-Weinberg equilibrium for haplotypes, penetrance model, and disease prevalence. We conducted sensitivity analyses to examine the robustness of ROB-HAP with respect to each of these model assumptions.

First, we assessed the robustness of ROB-HAP with respect to Hardy-Weinberg equilibrium. We generated data under model 
\[
\pr(h_{i}h_{j})= \left\{\begin{array}{cl}
F \mu_i+(1-F) \mu_i^{2}, & \text { for } i=j ,\\
2(1-F) \mu_i \mu_j, & \text { for } i \neq j,
\end{array}\right.
\]
where $F$ is a fixation index and $h_{i}h_{j}$ is an unordered diplotype. As shown in Table S3 (Supplementary Material), ROB-HAP had well controlled type-I error rates even when Hardy-Weinberg equilibrium was seriously violated ($F>0.5$). This was not surprising since Hardy-Weinberg equilibrium is required only in the ratio of $\pr(h_{iu}^{m}h_{ju}^{m},h_{wu}^{c},h_{lu}^{c})$ to $\pr(g^m_u)$ in the modified profile likelihood \eqref{equ:modified}, which is partially free of the fixation index $F$. In fact, when $g_u^m = 1$, we have
\[
\frac{\pr(h_{iu}^{m}h_{ju}^{m},h_{wu}^{c},h_{lu}^{c})}{ \pr(g_u^m)} = \frac{(1-F)\mu_i \mu_j \mu_l }{2(1-F)\theta (1-\theta)} = \frac{\mu_i \mu_j \mu_l }{2\theta (1-\theta)},
\]
which is free of the fixation index $F$; when $g_u^m \neq 1$, the parental origins can be directly derived from the genotypes of each mother-child pair. Therefore, the modified profile likelihood function \eqref{equ:modified} is only weakly dependent of $F$.

Second, we studied the impact of misspecifying the penetrance model on the performance of ROB-HAP. Data were generated under the penetrance model \eqref{equ:penetrance} with $\beta_{g^m X} = \beta_{g^c X} = \log1.2$ or $\beta_{g^m X} = \beta_{g^c X} = \log1.5$. The resultant underlying model had an interaction effect, but it was ignored in ROB-HAP. By doing so, we could assess whether or not ROB-HAP is sensitive to the presence of interaction effects. As shown in Table \ref{tab4}, the estimation of $\beta_{g^m}$, $\beta_{g^c}$, and $\beta_{X}$ were biased, but this was not the case for $\beta_{im}$. This indicates that ROB-HAP is robust to the penetrance model misspecification to some extent when assessing the parental origin effects.

\begin{table}
	\centering
	\caption{Estimation results of ROB-HAP with the penetrance model being misclassified $(f = 0.01, \eta = \log3)$}\label{tab4}
	\begin{threeparttable}
        \begin{tabular*}{\hsize}{@{}@{\extracolsep{\fill}}ccccccccccc@{}} 
			\hline
			&       & \multicolumn{4}{c}{$\beta_{g^mX}=\beta_{g^cX}=\log1.2$$^a$} &  & \multicolumn{4}{c}{$\beta_{g^mX}=\beta_{g^cX}=\log1.5$$^a$} \\ \cline{3-6} \cline{8-11} 
			Log-OR        & True$^b$  & Bias$^c$          & SE $^d$         & SEE $^e$        & CP$^f$          &  & Bias$^c$         & SE $^d$          & SEE$^e$         & CP$^f$          \\ \hline
			$\beta_{g^m}$ & 0.588 & 0.143         & 0.298       & 0.298       & 0.931       &  & 0.418        & 0.386        & 0.387       & 0.854       \\
			$\beta_{g^c}$ & 0.405 & 0.245         & 0.187       & 0.185       & 0.744       &  & 0.699        & 0.225        & 0.223       & 0.100       \\
			$\beta_{im}$  & 0.405 & -0.012        & 0.191       & 0.197       & 0.957       &  & 0.001        & 0.231        & 0.239       & 0.960       \\
			$\beta_X$     & 0.588 & 0.272         & 0.150       & 0.150       & 0.576       &  & 0.710        & 0.201        & 0.203       & 0.024       \\ \hline
		\end{tabular*}
	\begin{tablenotes}
\item $^a$With a nonzero interaction effect but ignored in ROB-HAP fitted model; $^b$true parameter value; $^c$difference between the mean estimate and true parameter value; $^d$empirical standard error; $^e$mean estimated standard error; $^f$empirical coverage probability of the 95$\%$ confidence interval. 
	\end{tablenotes} 
\end{threeparttable}
\end{table}

Finally, we assessed the robustness of ROB-HAP with respect to the disease prevalence $f$. We generated data with $f=0.01$ but misspecified it to be 0.002 or 0.05. The estimation results shown in Table \ref{tab5} demonstrate that the misspecification had little impact on the odds-ratio parameters of interest. Such robustness was also observed previously \citep{zhang2020efficient}.

\begin{table}
	\centering
	\caption{Estimation results of ROB-HAP with the prevalence being misclassified $(f = 0.01, \eta = \log3)$}\label{tab5}
	\begin{threeparttable}
        \begin{tabular*}{\hsize}{@{}@{\extracolsep{\fill}}ccccccccccc@{}}
			\hline
			&       & \multicolumn{4}{c}{Misspecified $f = 0.002^a$} &  & \multicolumn{4}{c}{Misspecified $f = 0.05$$^a$} \\ \cline{3-6} \cline{8-11} 
			Log-OR        & True$^b$  & Bias$^c$   & SE$^d$    & SEE$^e$   & CP$^f$    &  & Bias$^c$   & SE$^d$    & SEE$^e$   & CP$^f$    \\ \hline
			$\beta_{g^m}$ & 0.588 & 0.023  & 0.260 & 0.260 & 0.946 &  & -0.009 & 0.270 & 0.267 & 0.945 \\
			$\beta_{g^c}$ & 0.405 & -0.005  & 0.180 & 0.170 & 0.934 &  & 0.023  & 0.187 & 0.187 & 0.952 \\
			$\beta_{im}$  & 0.405 & -0.011  & 0.177 & 0.183 & 0.954 &  & 0.016  & 0.205 & 0.203 & 0.950 \\
			$\beta_X$     & 0.588 & 0.011 & 0.125 & 0.125 & 0.952 &  & 0.001  & 0.123 & 0.125 & 0.957 \\ \hline
		\end{tabular*}
		\begin{tablenotes}
        \item $^a$Misspecified prevalence in ROB-HAP; $^b$true parameter value; $^c$difference between the mean estimate and true parameter value; $^d$empirical standard error; $^e$mean estimated standard error; $^f$empirical coverage probability of the 95$\%$ confidence intervals.
		\end{tablenotes} 
	\end{threeparttable}
\end{table}

\subsection{Handling missing child genotypes}\label{sim:missing}

In real data analyses, some individuals could be subject to missing genotypes. In our method ROB-HAP, the children missing genotypes can be naturally handled through an expectation-maximization algorithm. A na\"{i}ve approach to handling missing genotypes is to simply ignore those mother-child pairs with missing genotype at any SNP. The efficiency loss of this na\"{i}ve approach could be considerable if the number of SNPs is large. A small-scale simulation study was conducted to demonstrate the efficiency improvement of incorporating mother-child pairs with missing child genotypes. Simulation data were generated in the same way as in Section \ref{GENdata}, and the genotype of each SNP was then randomly dropped with a probability of 0.01. Three versions of ROB-HAP were applied to the simulated data. The first version (ROB-HAP-com) used complete genotypes, the second version (ROB-HAP-EM) incorporated missing genotypes through the expectation-maximization algorithm, and the third version (ROB-HAP-del) simply deleted those mother-child pairs with missing genotypes.  Estimation results for parental origin effects by the three versions are graphically displayed in Fig. S1 (Supplementary Material). The three versions were shown to be virtually unbiased. ROB-HAP-EM and ROB-HAP-com were nearly comparable, and they had smaller standard errors than ROB-HAP-del. These results evidently demonstrate that incorporating mother-child pairs with missing data can significantly improve the estimation efficiency.

\section{Application to the Jerusalem Perinatal Study}
\label{s:apply}
We applied the considered methods for assessing the parental origin effects in the gene \emph{PPARGC1A} on low birth weight in the Jerusalem Perinatal Study, a historical birth cohort study including 17,003 births in Jerusalem residents from 1974 to 1976. Among 8,238 eligible children with mothers' prepregnancy body mass index (pp-BMI) smaller than 25, there were 297 cases of low birth weight (below 2.5 kg). Genotype data were obtained for 691 eligible mother-child pairs (125 case pairs and 566 control pairs). We incorporated pp-BMI in applicable methods as a covariate because it was shown to be strongly associated with infant nutrition during pregnancy. The low birth weight prevalence $f$ was set to be the proportion of low birth weight children in the original data, i.e., 0.036 ($=297/8,238$). 

The genotypes on 24 SNPs in gene \emph{PPARGC1A} were available, five of which (i.e., rs3774921, rs3755863, rs8192678, rs2970849, and rs1472095) showed strong linkage disequilibrium using the \texttt{R} package \texttt{LDheatmap} based on allelic correlation $r^2$ and the physical distance between the SNPs. Exploratory analyses of the five candidate SNPs can be found in Table S9 of \cite{zhang2021covariate}. It was shown previously that SNP rs8192678 may have a parental origin effect on birth weight \citep{lin2013multi} when pp-BMI was not adjusted for. We therefore applied eight methods (i.e., ROB-HAP, ROB-SNP, LOGIT-HAP, LOGIT-SNP, IND-HAP, IND-SNP, P-HAP, and P-SNP) to re-assess parental origin effect of SNP rs8192678 on low birth weight adjusting for pp-BMI. Methods imposing independence assumption between rs8192678 and pp-BMI (i.e., IND-HAP and IND-SNP) are applicable since this SNP was not significantly associated with pp-BMI \citep{zhang2020efficient}. Among the 691 mother-child pairs, 15 had incomplete genotype data and 58 had incompatible genotypes at any of the five SNPs. Those child genotypes incompatible with their mothers' genotypes were treated as missing. Among the $691-15-58=618$ families with incomplete or incompatible genotypes, 145 had heterozygous genotypes at the SNP rs8192678 for both mothers and children. Among the 145 families, the parental origins of 73 families were unambiguously determined using the joint genotypes at the five SNPs. It happened that all of the children in these 73 families inherited the minor allele from their mothers.
Both ROB-HAP and IND-HAP could accommodate missing child genotypes via our proposed expectation-maximization algorithm. The other methods only used those families with complete genotypes.

\begin{table}
	\centering
	\caption{Parental origin effect estimates for SNP rs8192678 in the Jerusalem Perinatal Study}\label{tab6}
	\begin{threeparttable}
        \begin{tabular*}{\hsize}{@{}@{\extracolsep{\fill}}ccccc@{}}
			\hline
			Method$^a$      & EST$^b$     & SE$^c$      & 95\% CI$^d$ & P-value$^e$   \\ \hline
			IND-SNP     & 0.612 & 0.256 & {[}0.111, 1.114{]}  & 0.017   \\
			ROB-SNP     & 0.597 & 0.258 & {[}0.092, 1.102{]}  & 0.020   \\
			LOGIT-SNP   & 0.827 & 0.373 & {[}0.096, 1.558{]}  & 0.027   \\
			P-SNP       & 0.611 & 0.256 & {[}0.110, 1.112{]}  & 0.017   \\
			IND-HAP     & 0.368 & 0.236 & {[}-0.096, 0.831{]} & 0.120   \\
			IND-HAP$^f$   & 0.320 & 0.226 & {[}-0.122, 0.763{]} & 0.156   \\
			ROB-HAP     & 0.340 & 0.238 & {[}-0.127, 0.807{]} & 0.154   \\
			ROB-HAP$^f$   & 0.305 & 0.230 & {[}-0.145, 0.755{]} & 0.184   \\
			LOGIT-HAP   & 0.415 & 0.247 & {[}-0.069, 0.899{]} & 0.093   \\
			P-HAP       & 0.363 & 0.236 & {[}-0.100, 0.826{]} & 0.124   \\ \hline
		\end{tabular*}
		\begin{tablenotes}
\item $^a$Various methods mentioned in Section \ref{s:simu}; $^b$estimated parental origin effects; $^c$estimated standard errors for parental origin effects; $^d$95$\%$ Confidence intervals for parental origin effects; $^e$p-values for testing parental origin effects; $^f$families with missing data used.
		\end{tablenotes} 
	\end{threeparttable}
\end{table}

Parental origin effect estimates are summarized in Table \ref{tab6}. All methods produced positive parental origin effect estimates, indicating the minor allele (T) of the child with low birth weight was preferentially transmitted from the mother, though the estimated effect sizes varied to some extent. The estimation differences between the HAP and SNP methods could be due to the fact that all of the children in the 73 families with hetorozygous genotypes at the SNP rs8192678 inherited the minor allele from their mothers. The prospective likelihood methods LOGIT-SNP and LOGIT-HAP were uniformly less efficient than the retrospective likelihood based methods. For example, the SE by LOGIT-SNP was 0.373, compared with 0.256, 0.258, and 0.256 by IND-SNP, ROB-SNP, and P-SNP. This verified the simulation results in Section \ref{simu1}. 
Furthermore, utilizing haplotypes could effectively improve the estimation efficiency for the parental origin effects. In fact, the SEs were 0.256, 0.258, 0.383, and 0.256 by IND-SNP, ROB-SNP, LOGIT-SNP, and P-SNP, compared with 0.236, 0.238, 0.247, and 0.236 by IND-HAP, ROB-HAP, LOGIT-HAP, and P-HAP. This result was consistent with the simulation results in Section \ref{simu2}. Furthermore, utilizing families with missing or incompatible genotypes slightly improved estimation efficiency, that is, the SE was reduced from 0.238 to 0.230 with ROB-HAP and from 0.236 to 0.226 with IND-HAP. Finally, P-HAP and ROB-HAP had comparable SEs, which was consistent with the simulation results in Section \ref{simu3}.
\section{Discussion}
\label{s:discuss}

Testing parental origin effects is an important genetic problem, and it is a major challenge to develop an inference procedure of high efficiency utilizing case-control mother-child pair data. Under this study design, we develop a modified retrospective-likelihood-based statistically efficient inference method, ROB-HAP, for testing parental origin effects while controlling for the confounding effects of covariates. Genotypes from genetic markers tightly linked with the target genetic marker are used to accurately infer the parental origin. ROB-HAP can fully exploit genetic information in the available data, including Hardy-Weinberg equilibrium, Mendelian law between maternal genotype and child genotype, and the conditional independence between covariates and child genotype given maternal genotype. ROB-HAP is statistically robust to some extent in the sense that no parametric assumption is imposed on the relationship between the maternal genotypes and covariates. A modified profile likelihood function is adopted to ensure computational robustness, without sacrificing statistical efficiency. Simulation results demonstrated that ROB-HAP was very robust to the violation of Hardy-Weinberg equilibrium and misspecification of disease prevalence. It was also shown to be robust to the presence of unaccounted gene-environment interaction effects in testing parental origin effects.

The proposed method ROB-HAP can be improved from several aspects.
First, ROB-HAP assumes an additive mode of inheritance for the genetic effects, but this assumption can be straightforwardly relaxed by introducing dominance effects. Second, the robustness of ROB-HAP and the effciency of IND-HAP can be balanced by adopting a shrinkage estimator similar to the one proposed in \cite{chen2009shrinkage}. Specifically,
let $\hat{\bm{\beta}}_\textrm{ROB}$ and $\hat{\bm{\beta}}_\textrm{IND}$ denote the estimators of regression parameter vector $\bm{\beta}$ using ROB-HAP and IND-HAP, respectively, and let $\beta_\textrm{ROB}$ and $\beta_\textrm{IND}$ denote their limiting values. The parameter $\psi = \beta_\textrm{IND} - \beta_\textrm{ROB}$ can be used to describe the degree of deviation between the true model and the independence model. A shrinkage estimator $\hat{\bm{\beta}}_\textrm{EB}$ can be derived under the empirical Bayesian framework: $\hat{\bm{\beta}}_\textrm{EB} = \hat{\bm{\beta}}_\textrm{ROB} + K(\hat{\bm{\beta}}_\textrm{IND} - \hat{\bm{\beta}}_\textrm{ROB})$, where $K = V(V + \hat{\psi}\hat{\psi}^\mathrm{\scriptscriptstyle T})^{-1}$, $\hat\psi=\hat{\bm{\beta}}_\textrm{IND} - \hat{\bm{\beta}}_\textrm{ROB}$, and $V$ is an appropriate consistent estimator of $\textrm{cov}(\hat{\psi})$. Third, the proposed method can be extended to  family data of other types. In some studies, genotype data can be obtained from some mother-father-child triads. The paternal genotype is helpful in determining the parental origin, thus improving statistical inference efficiency. ROB-HAP can be immediately extended to handle such families by adding paternal genotypes in the retrospective likelihood \eqref{equ:retro}. Statistical efficiency can be further improved when genotypes are available from multiple children in some families. Various types of family structures may be present in real studies, and it deserves further investigation to combine these data.

In practice, we recommend using four adjacent SNPs in ROB-HAP and ignoring haplotypes with small estimated frequencies (e.g. less than 0.01) since involving more adjacent SNPs has very limited power gain in testing parental origin effects but significantly increase computational burden. In some studies, the collected families may have multiple ethnicities. If the ancestry information is known for each family, a stratified analysis can be easily conducted by utilizing Fisher's P-value combination method \citep{fisher1992statistical,ray2016usat}; otherwise, principle components extracted from genomewide data can be used to adjust for population structure.

\section*{Acknowledgement}
The work of HZ was partially supported by the Natural Science Foundation of China (No. 12171451), and the work of JC was partially suported by the grants from the National Institutes of Health (No. R21-ES020811 and R01-ES016626).

\section*{Supplementary material}
\label{SM}
Supplementary material 
includes derivations or proofs of the theoretical results and additional simulation results. (https://github.com/yatian20/APOE).

\bibliographystyle{apalike}
\bibliography{ref}
\end{document}